\documentclass[twocolumn,floatfix]{revtex4}
\usepackage{graphicx}
\usepackage{dcolumn}
\usepackage{amsmath}

\usepackage{amsmath,amsfonts,amssymb}
\usepackage{wrapfig}
\usepackage{graphicx}
\usepackage{bbm}
\usepackage{graphics}

\def \beq {\begin{equation}}
\def \eeq {\end{equation}} 
\def \ba {\begin{eqnarray}}
\def \ea {\end{eqnarray}}

\usepackage{amsmath,amsfonts,amssymb}
\usepackage{wrapfig}
\usepackage{graphicx}
\usepackage{bbm}
\usepackage{graphics}

\def \beq {\begin{equation}}
\def \eeq {\end{equation}}
\def \ba {\begin{eqnarray}}
\def \ea {\end{eqnarray}}

\newcommand{\Bo}{{B}_{\rm 0}}
\newcommand{\Btot}{{\bf{B}}_{\rm tot}}

\newcommand{\Btotz}{B\mathrm{_{tot}}}

\begin{document}
\hsize\textwidth\columnwidth\hsize\csname@twocolumnfalse
\endcsname

\title{Exchange Control of Nuclear Spin Diffusion in a Double Quantum Dot}
\author{D. J. Reilly$^{1}$, J. M. Taylor$^{2}$, J. R. Petta$^3$, C. M. Marcus$^1$, M. P. Hanson$^4$ and A. C. Gossard$^4$}
\affiliation{$^1$ Department of Physics, Harvard University, Cambridge, MA 02138, USA}
\affiliation{$^2$ Department of Physics, Massachusetts Institute of Technology, Cambridge, MA 02139, USA}
\affiliation{$^3$ Department of Physics, Princeton University, Princeton, NJ 08544, USA}
\affiliation{$^4$ Department of Materials, University of California, Santa Barbara, California 93106, USA} 

\maketitle

{\bf Coherent two-level systems, or qubits, based on electron spins in GaAs quantum dots are strongly coupled
  to the nuclear spins of the host lattice via the
  hyperfine interaction \cite{Petta_science05, Erlingsson,Khaetskii, Merkulov, Witzel_PRB06}.
  Realizing nuclear spin control would likely improve electron spin coherence
  and potentially enable the nuclear environment to be harnessed for the
  long-term storage of quantum information
  \cite{Taylor_PRL03,Witzel_PRB07}. Toward this goal,
  we report experimental control of the relaxation of nuclear spin
  polarization in a gate-defined two-electron GaAs double quantum dot.  A cyclic
  gate-pulse sequence transfers the spin of an electron pair to the host nuclear
  system, establishing a local nuclear polarization that
  relaxes on a time scale of seconds.  We find nuclear relaxation depends on magnetic field and gate-controlled two-electron exchange, consistent with a model of electron-mediated nuclear spin diffusion.}

\begin{figure}[t!!]
\begin{center}
\includegraphics[width=8.5cm]{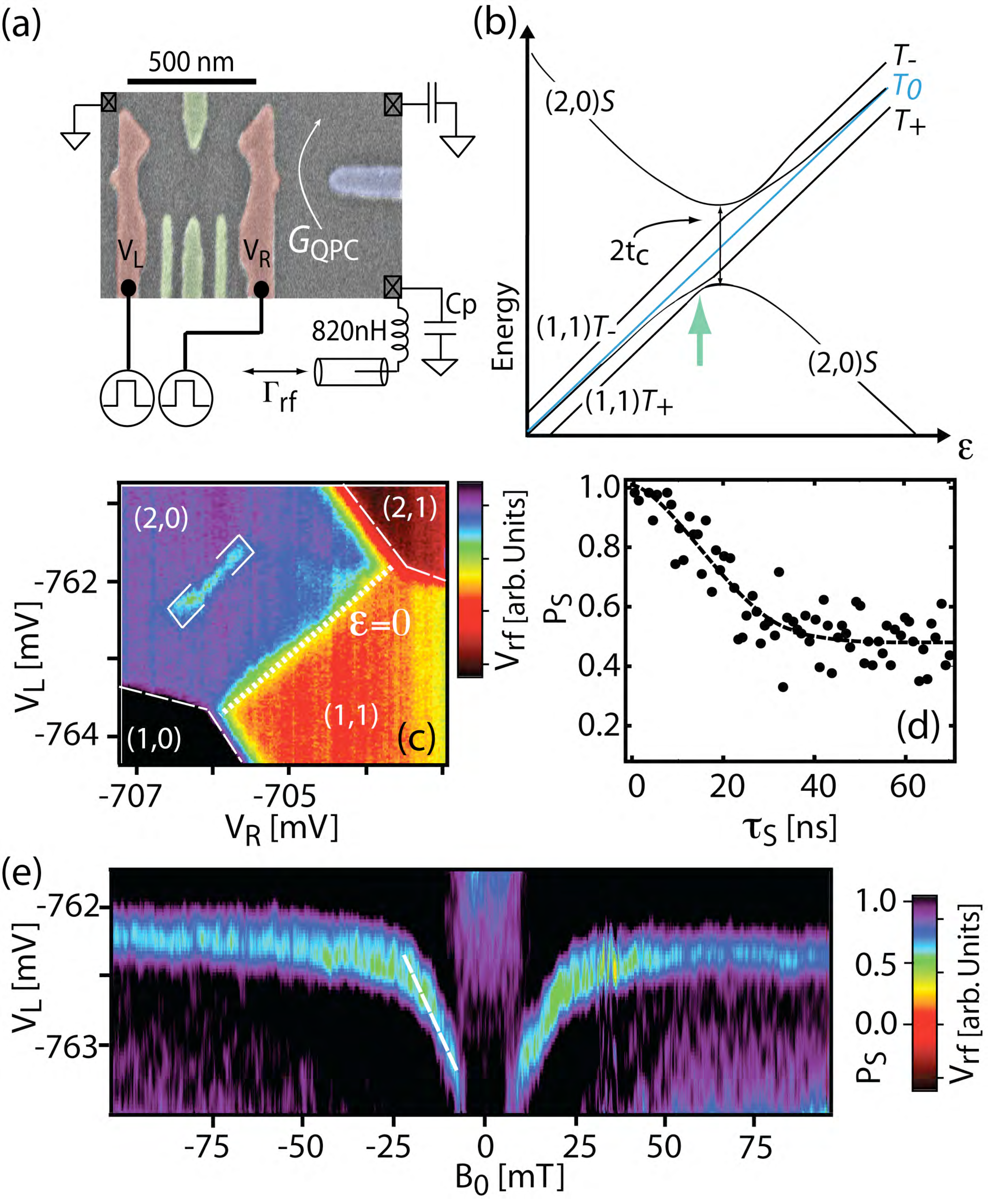}
\caption{(a) False-color SEM image of a representative
  double-dot with integrated rf-QPC charge sensor. Changes in QPC
  conductance are mapped to changes in reflected rf power from an
  impedance matching network based on series inductor (820 nH) and
  paracitic capacitance $C_\mathrm{p}\sim$ 0.6 pF. (b) Energy level
  diagram of the two-electron system. Labels (n,m) define the number
  of electrons in the left and right dot. Green arrow points to the $S-T_{+}$ avoided crossing.  (c) $V_\mathrm{rf}$ around the (2,0)-(1,1) charge transition during cycling of the probe sequence. A plane has been subtracted. The region indicated with white lines corresponds to the $S-T_{+}$
  resonance. $\Bo$ = 8 mT. (d) Singlet return probability $P_{S}$ as a
  function of separation time  $\tau_\mathrm{S}$ yielding a $T_{2}^{*}\sim$ 15
  ns.  $\Bo$ = 8 mT, $\tau_{M}$ = 1.6 $\mu$s. Black dashed line is a fit to the theoretical gaussian
  form \cite{taylor_theory}. (e) $P_{S}$ as a function of left gate
  bias $V_{L}$ and magnetic field $\Bo$. Dashed white line is used to
  convert position of the resonance in $V_{L}$ to $\Btotz$.}
\vspace{-0.5cm}
\end{center}
\end{figure}

Dynamic nuclear polarization (DNP), in which the `flip' of a polarized
electron spin is accompanied by the simultaneous `flop' of a nuclear
spin \cite{Abragam}, has served as a probe of nuclear dynamics in bulk
semiconductors \cite{Lampel_PRL68, Paget_PRB82}, confined semiconductor devices,
and optical systems
\cite{Wald_PRL94,Dobers_PRL_61,Salis_PRL01,Gammon_PRL01,Maletinsky_PRL07,Lai_PRL06,Tartakovskii_PRL07}. In quantum dots, hyperfine coupling with electron spins can lead to
nuclear dynamics distinct from those of bulk materials. For instance, using optical techniques, the presence of a single residual electron in an InGaAs dot was recently shown to significantly enhance the decay of nuclear polarization \cite{Maletinsky_PRL07}.  Signatures of DNP have also been investigated in transport through few-electron
double quantum dots. In this case, spin blockade can lead to a complex
interplay between electron and nuclear spin transitions, resulting in
bi-stability, hysteresis, and long-time oscillations of leakage
currents  \cite{Ono_PRL04,Koppens_science05,Baugh_PRL07,Nazarov,Rudner}. 
Understanding the coupled evolution of the electron-nuclear system 
is important for the development of long-lived qubits based on these devices.

In this Letter, we report time-resolved measurements investigating the induction and relaxation of DNP
in a few-electron double quantum dot as a function of magnetic field and charge arrangement. Cyclic evolution of the two-electron spin state, driven by gate pulses \cite{Petta_DNP}, repeatedly flops nuclear spins to create a small local DNP of order 1\%. Relaxation is monitored by detecting the Overhauser field using high-bandwidth proximal charge sensing \cite{Reilly_APL07}. From the long nuclear relaxation times we conclude that the modest polarization achieved is not limited by nuclear spin out-diffusion, but rather likely arise from a saturation in the flip-flop efficiency of the pumping cycle. The present work advances previous studies by demonstrating that nuclear diffusion can be made sensitive to the exchange coupling of confined electrons, controlled experimentally through the spatial charge arrangement with fixed total charge. Finally, we infer from magnetic field and charge-arrangement dependences that electron-mediated coupling of nuclear spins \cite{Abragam,Coish_Loss05,Sham_PRB06} is the dominant contribution to the nuclear polarization diffusion rate.
 
\begin{figure}[t!!]
\begin{center}
\includegraphics[width=8.0cm]{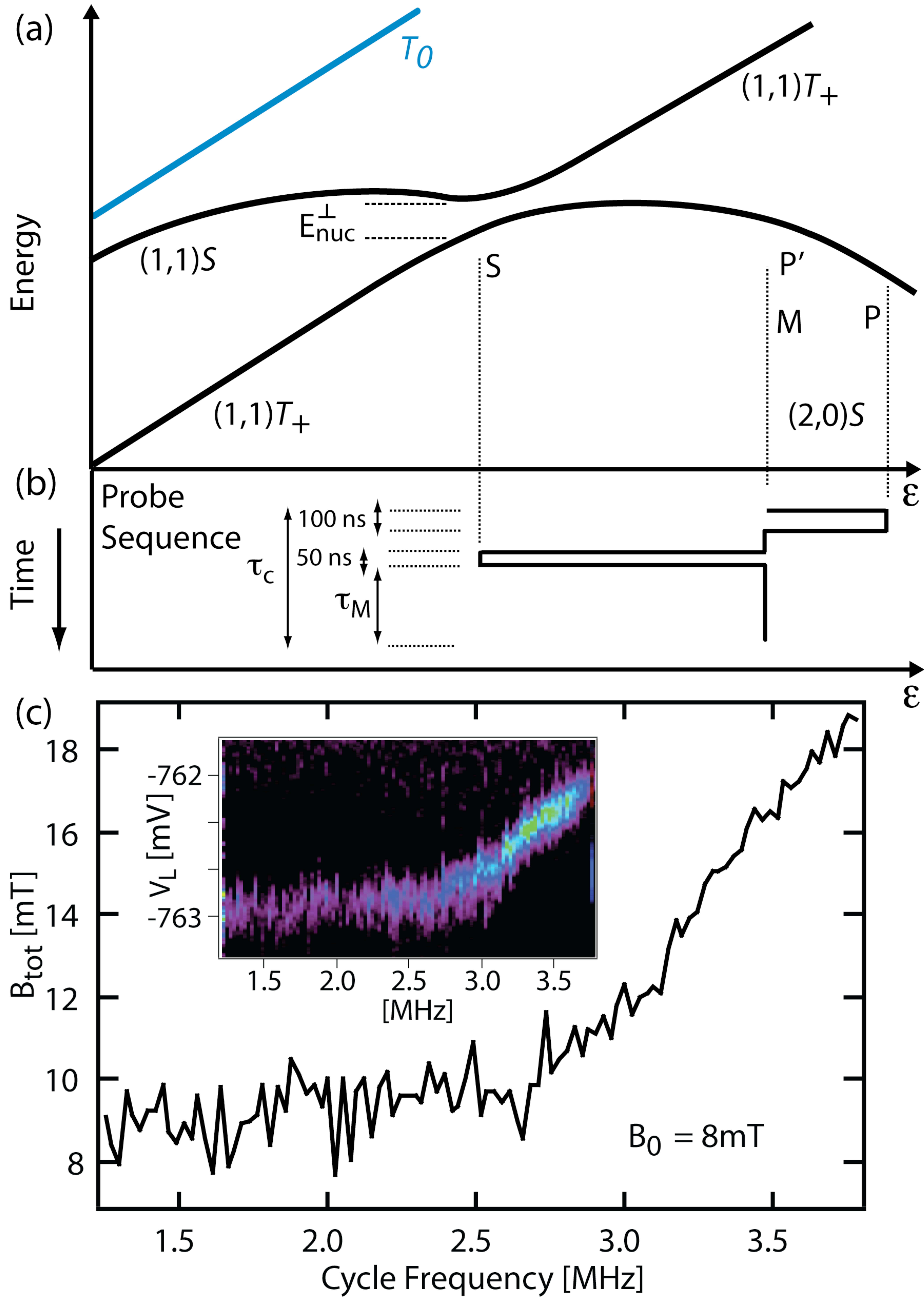}
\caption{(a) Energy level diagram near the $S-T_{+}$
  resonance. (b) Pulse cycle used to measure the position of  the resonance during the ``probe" sequence. (c)
  Inset: Position of $S-T_{+}$ resonance with respect to gate bias,
  $V_{L}$. Color scale is the same as Fig.~1(e). For cycle rates
  below 1 MHz, the position of the resonance indicates $\Btotz
  \sim \Bo$, i.e., no appreciable polarization is established by the process of measuring the position of the $S-T_{+}$ resonance.  The main panel shows the position of the resonance converted to units of magnetic field via the calibration in Fig.~1(e).}
\vspace{-0.5cm}
\end{center}
\end{figure}

The double quantum dot is formed by Ti/Au gates patterned with
electron beam lithography on the surface of a
GaAs/Al$_{0.3}$Ga$_{0.7}$As heterostructure with two dimensional
electron gas (2DEG) with density 2$\times$10$^{15}$ m$^{-2}$ and mobility 20
m$^{2}$/Vs, as shown in Fig.~1(a). Measurements were made in a dilution
refrigerator at the base electron temperature of $\sim$ 120 mK. A
schematic energy level diagram of the two-electron system is shown in
Fig.~1(b), with the labels (n,m) giving the number of electrons in the
left and right dot. Quasistatic gate voltages control interdot tunnel coupling, 
$t_{c}$, while the detuning, $\epsilon$, from the (2,0)-(1,1) charge degeneracy
is controlled on nanosecond time scales using fast pulses applied to the gates marked in red in Fig.~1(a). The charge configuration of the double dot is 
detected by monitoring the conductance $G_\mathrm{QPC}$ of a proximal
rf quantum point contact (rf-QPC). $G_\mathrm{QPC}$ controls the reflected power $\Gamma_\mathrm{rf}$ of a 220 MHz rf-carrier; following demodulation, this yields a voltage
 $V_\mathrm{rf}$ that constitutes the charge sensing signal \cite{Reilly_APL07}.

The effective total field experienced by electrons in (1,1) is given by $\Btot = {\bf{\Bo}} + {\bf{B}}_\mathrm{nuc}$, where $\bf{\Bo}$ is the external field applied perpendicular to the 2DEG plane and ${\bf{B}}_\mathrm{nuc} = ({\bf{B}}_\mathrm{nuc}^\mathrm{L} + {\bf{B}}_\mathrm{nuc}^\mathrm{R})/2$ is the Overhauser field averaged over left and right dots.  The avoided crossing between the singlet ($S$) and the (1,1) $m_s =1$ triplet ($T_+$) occurs at a value of $\epsilon$ (green arrow in Fig.~1(b)) set by the total Zeeman energy, $E_\mathrm{{tot}} = g\mu_{B} \Btotz$, where $g \simeq -0.4$ is the electron g-factor in GaAs, $\mu_{B}$ is the Bohr magneton, and
$\Btotz$ is the magnitude of $\Btot$. The gap and width of the avoided crossing are set by $E^{\perp}_\mathrm{nuc} = g \mu_{B} \Delta B_\mathrm{nuc}^{\perp}$, where $\Delta B_\mathrm{nuc}^{\perp}$ is the magnitude of the component of $\Delta {\bf{B}}_\mathrm{nuc}  =  ({\bf{B}}_\mathrm{nuc}^\mathrm{L} - {\bf{B}}_\mathrm{nuc}^\mathrm{R})/2$ transverse to $\Btotz$.

We probe the $S-T_{+}$ resonance using the pulse sequence shown in
Fig.~2(b), which first prepares (2,0)$S$ at (P) then separates the
electrons (S) for a time  $\tau_\mathrm{S}$ before returning to (2,0) for
measurement (M) for time $\tau_\mathrm{M} \sim$ 5~$\mu$s. Pauli spin-blockade ensures that only the (1,1) singlet returns to (2,0), with triplets blocked for a time $T_{1}$. In this way,  the two-electron spin-state is mapped to a charge configuration that is detected with the rf-QPC. Cycling this sequence
yields a feature at (M) in the (2,0) region, indicated by white lines
in Fig.~1(c). Once calibrated, $V_\mathrm{rf}$ gives the
probability 1-$P_{S}$ that an initial singlet evolved into $T_{+}$ during the separation interval $\tau_\mathrm{S}$.   $V_\mathrm{rf}$ is calibrated using the measured values in (2,0) and (1,1) to define $P_{S}$ = 1
and $P_{S}$ = 0, giving the scale bar in
Fig.~1(e).  Fitting the time-averaged function $P_{S}(\tau_\mathrm{S})$ gives an inhomogeneous dephasing time,  $T^{*}_{2} \sim$~15~ns. The dependence of the $S-T_{+}$ resonance position (in $V_{L}$, with $V_{R}$ fixed) on $B_0$ in the range  $B_0 = 5 - 18$ mT, in the absence of a time-averaged nuclear polarization, serves as a calibration used to determine $\Btotz$ when nuclear polarization is present \cite{foot_funnel}.

Dynamic nuclear polarization is investigated using a three-step ``pump-pause-probe" sequence: The pump sequence starts from a singlet in (2,0) then moves adiabatically through the $S-T_+$ resonance, flipping an electron and flopping a nuclear spin, in principle once per cycle at a rate of 4 MHz \cite{Petta_DNP}. The ``probe'' sequence (Fig.~2(a,b)) also starts with a singlet in (2,0) but moves to the $S-T_+$ resonance, providing a measure of $\Btotz$. For a cycle rate below 1 MHz, the probe sequence does not induce nuclear polarization, as seen in Fig.~2(c). For all DNP data shown, the cycle rate of the probe sequence was 200 kHz. Pump and probe cycles are separated by a static ``pause" of duration $\Delta t$.

The pump sequence creates a steady-state DNP of  order $\sim 10$ mT, which, in the absence of a pause,
relaxes during the probing cycle on a time scale $\tau_{R} =
8$ s, found by fitting an exponential to $\Btotz (t)$ (Fig.~3(c)).
Increasing $\Bo$ from 8~mT to 10~mT doubles the time
taken for $\Btotz$ to return to $\Bo$. At $\Bo$ = 15~mT,  $\Btotz$
relaxes over a time scale similar to the $\Bo$ = 10~mT data. 
 We note that at $t$ = 0, $\Btotz$ appears nearly independent of $\Bo$. This suggests that the pump sequence ceases to produce polarization above a certain value of $\Btotz$, consistent with previous measurements \cite{Petta_DNP}.  The measured relaxation rate cannot account for the small steady-state polarization ($\sim$ 10 mT), and we are led to conclude that there must be significant decrease in the efficiency of the polarization cycle with increasing $B_{\rm nuc}$.
\begin{figure}[t!!]
\begin{center}
\includegraphics[width=8.0cm]{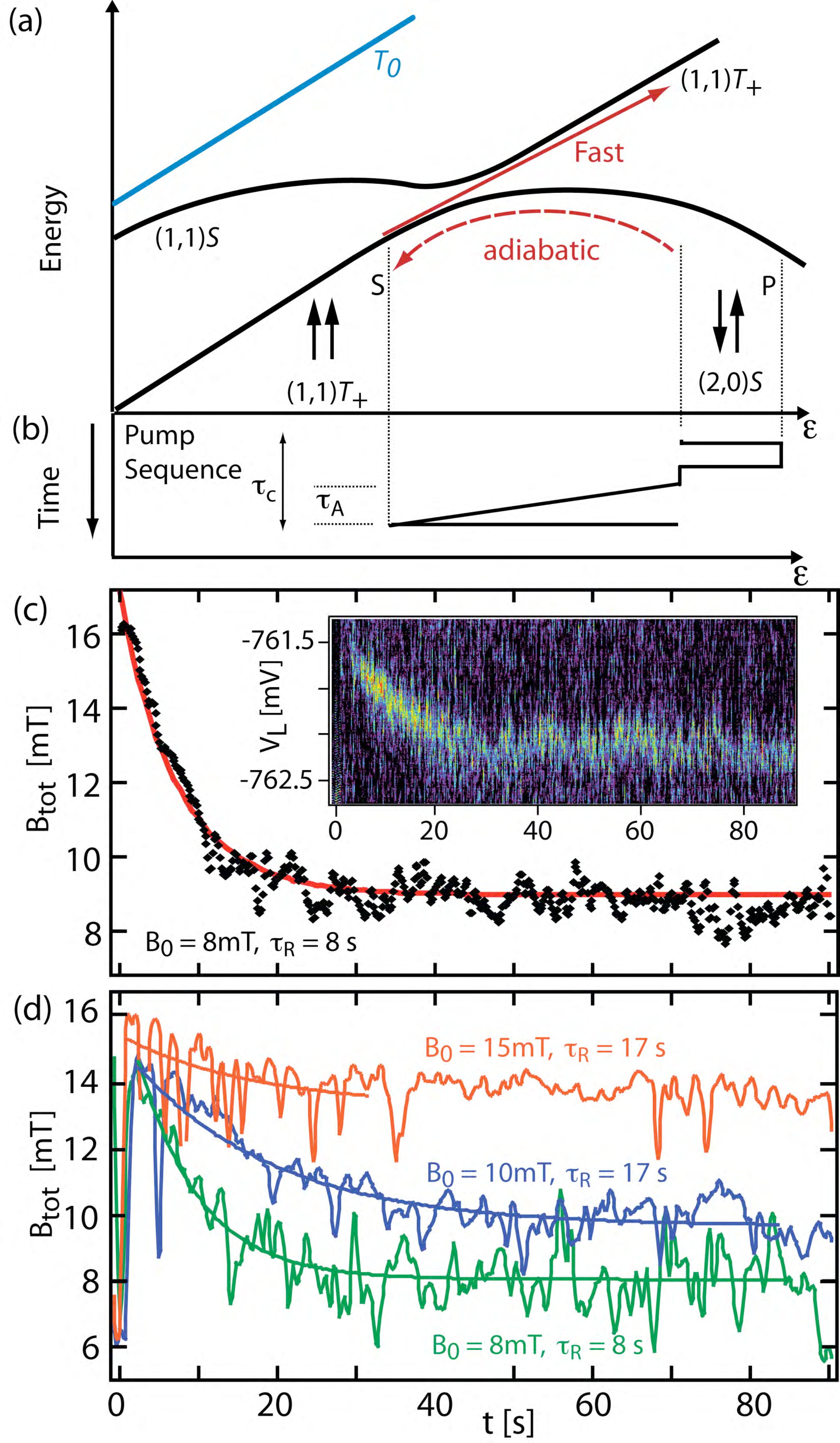}
\caption{(a) Energy level diagram near the $S-T_{+}$
  avoided-crossing with pump sequence used to create a DNP shown in
  (b), $\tau_{A}$ = 50 ns, $\tau_{c}$ = 250 ns. Pump cycle rate is 4 MHz. (c) Inset: Decay in the position of the resonance with respect  to $V_{L}$ following pumping. Main panel shows the average of five
  pump-probe sequences, with $\Btotz$ calibrated using
  Fig.~1(e). Red curve is an exponential fit. (d) Relaxation of DNP at $\Bo$ = 8 mT (green) $\tau_{R} = 8 \pm 2$ s, $\Bo$ =
  10 mT (blue) $\tau_{R} = 17 \pm 3$ s, and $\Bo$ = 15 mT (red) $\tau_{R} = 17 \pm 5$ s. For the $\Bo$ = 15 mT data we constrain the fit to the first 30 s due to the small available polarization signal. Noise in the resonance position exaggerates $\tau_{R}$ to 22 s when fitting over the total data range.}
\vspace{-0.5cm}
\end{center}
\end{figure}

The effect of pausing in (2,0) between the pump and probe sequences can be seen in Fig.~4(b), which shows that
more than half the polarization remains after pausing for 30~s in (2,0)S (Fig.~4(c)).  Once the probe sequence is initiated after the pause, $\Btotz$ once again
decays with $\tau_{R} \sim$ 8 s. The influence of the probe sequence is examined further by introducing
multiple pause intervals in (2,0), interleaved with probe cycles (Fig.~4(d)).

\begin{figure*}[t!!]
\begin{center}
\includegraphics[width=16cm]{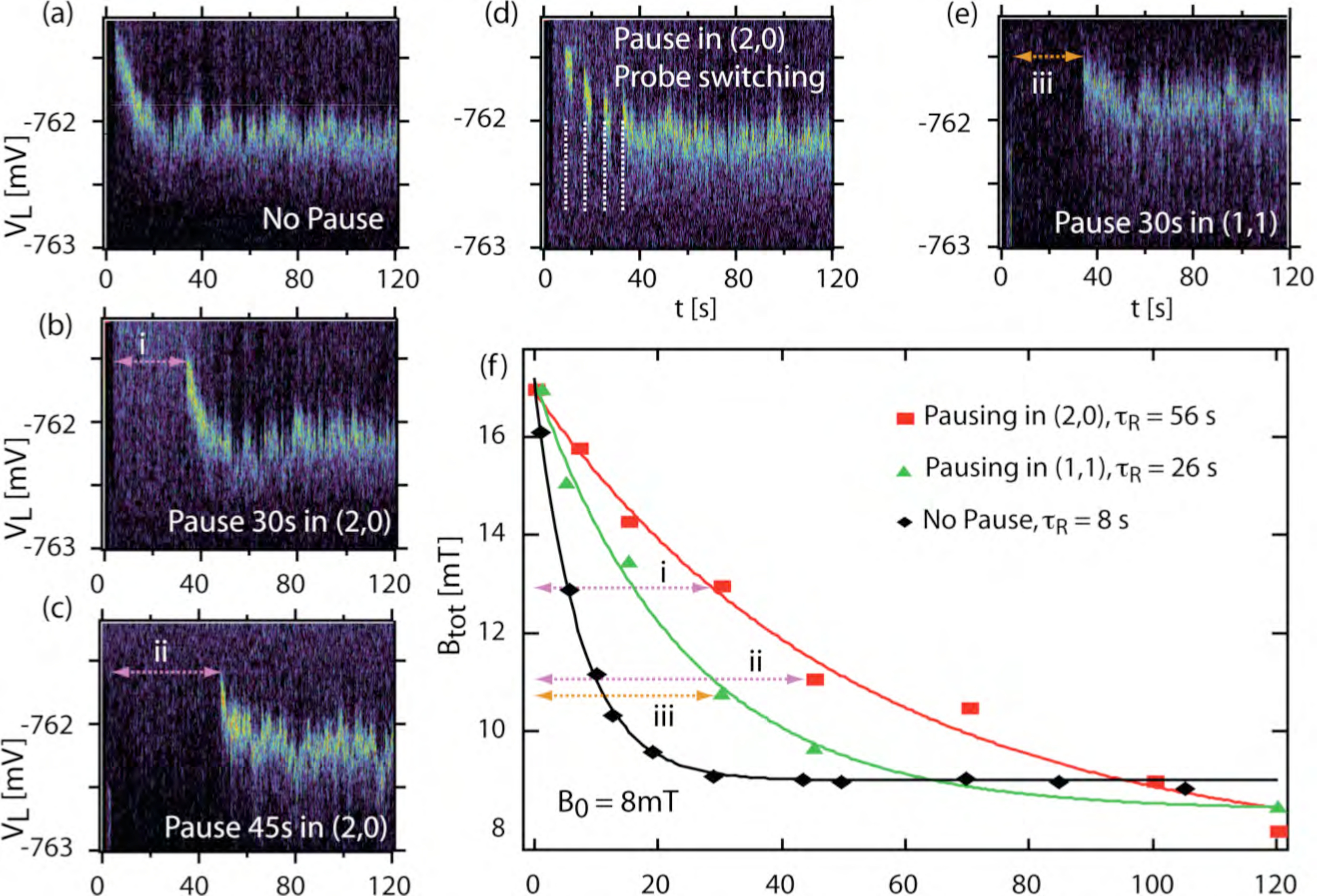}
\caption{(a) Immediate decay in the position of the
  resonance during the probe sequence. (b) Decay in position of the
  resonance following a pause interval of 30 s in (2,0) between the pump
  and probe sequences. Pausing in (2,0) suppresses hyperfine
  coupling. (c) Same as (b), but with pause interval set to 45 s. (d)
  Decay of resonance during probing, interleaved with multiple pause
  intervals. (e) Decay in position of resonance following a pause of
  30 s in (1,1). (f) Decay of $\Btotz$ as a function of
  pause interval $\Delta t$ and for different configurations of two-electron spin
  state.}
\vspace{-0.5cm}
\end{center}
\end{figure*}

The dependence of the nuclear relaxation rate on the two-electron spin-state during the pause duration is shown in Fig.~4(f). Pausing for the duration of $\Delta t$  in the (2,0) state yields a relaxation time $\tau_{R}$ = 56~s (red data in Fig.~4(f)), while pausing in (1,1) yields $\tau_{R}$ = 26~s (green data in Fig.~4(f)) \cite{foot_recal}. We ascribe these different relaxation times to a nuclear spin diffusion constant that depends
on the two-electron spin state.  With diffusion dominated by the shortest dimension of the dot, perpendicular to the electron gas, we approximate the diffusion constant $D= d^{2}/\tau_{R}$ based on an estimate of the width of the wavefunction $d \sim$ 7.5~nm \cite{Stopa}. This gives $D \sim 1\times 10^{-14}$ cm$^{2}$s$^{-1}$ for the case of pausing in (2,0), consistent with estimates of diffusion by nuclear dipole-dipole flipping \cite{Paget_PRB82}. Activation of the probe sequence increases diffusion to $D \sim 7 \times 10^{-14}$ cm$^{2}$s$^{-1}$.

The two-electron spin state is expected to affect nuclear spin diffusion in two opposing ways. The presence of strongly confined electrons creates an inhomogenous Knight shift \cite{Khaetskii}, lifting the degeneracy between nuclear dipoles and suppressing diffusion by dipolar flipping. Competing with this mechanism is the enhancement of diffusion via electron-mediated nuclear spin exchange \cite{Abragam}.  We first estimate the magnitude of each of these mechanisms in an effort to explain the different diffusion rates observed in (1,1) and (2,0). For our device we estimate that the Knight shift alone suppresses diffusion by at most 10 \% (see Supplementary Information). Indirect nuclear-spin exchange occurs when a nuclear spin, contributing to $\Delta {\bf{B}}_\mathrm{nuc}^\perp$,  flips the electron spin, which, upon flipping back generally flops a different nuclear spin  \cite{Coish_Loss05,Sham_PRB06}. This virtual process of nuclear-spin exchange is suppressed by the electron Zeeman energy and is thus dependent on $\Btotz$. We note that mediated flipping operates on nuclear spins within the dot, enhancing diffusion from the center to the edge, where dipolar diffusion beyond the dot begins to dominate. We find (see Supplementary Information) that for $\Btotz \lesssim 10~B_\mathrm{nuc} \sim$ 20~mT, the enhancement of diffusion via electron-mediated spin flips in the (1,1) state dominates the suppression of diffusion due to the Knight shift, leading to an overall increase in nuclear spin diffusion.  However, with electrons in (2,0)$S$, both hyperfine mechanisms are suppressed by the electron exchange energy $J$, which is $10^{4}$ times larger than $g \mu_B \Btotz$ for the fields used in these experiments. In particular, electron-mediated (enhanced) diffusion is a negligible contribution and the dynamics of $\Btotz$ are governed solely by the bare dipole-dipole diffusion of polarization from the dot \cite{McNeil_PRB76}.

Electron mediated flipping leads to an increase in diffusion with decreasing $\Btotz$, consistent with the $\Bo$ dependence of the data shown in Fig.~3(d).  Non-secular corrections to the nuclear dipole-dipole interaction will also enhance diffusion for $\Btotz \lesssim 1$ mT \cite{Abragam,Maletinsky_PRL07}, but these are suppressed at the applied fields used in our experiment. Flipping of spins via co-tunneling is estimated to be a negligible based on a measurement of electron spin relaxation ($T_1 \sim 15 \mu$s) in this device. At the $S-T_{+}$ resonance, exchange and the electron Zeeman energy effectively ``cancel'' allowing rapid flipping of electrons that readily mediate rapid exchange of nuclear spins. This is the likely explanation for the enhanced diffusion observed during the probe sequence. 

Hyperfine mediated nuclear dynamics in quantum dots have been considered theoretically in the context of spin-preserving processes \cite{Erlingsson,Merkulov,Witzel_PRB06,Khaetskii,Coish_Loss05,Sham_PRB06}, but measurement of the time scales for nuclear relaxation in dots containing a single electron have only recently been reported \cite{Maletinsky_PRL07}. For two-electron systems, the measurements presented here bring to light the 
role of electron exchange, which as we have shown can lead to a suppression of hyperfine-mediated nuclear spin diffusion. Finally, based on our measurement of $\tau_{R}$, we emphasize that the maximum steady-state DNP $\sim$ 10 mT cannot be limited by rapid out diffusion. Rather, these results indicate that the pump sequence strongly decreases in efficiency with increasing polarization, consistent with previous measurements \cite{Petta_DNP} and the idea of dark state formation \cite{Imamoglu_PRL03}. We anticipate that these results will be of relevance in the construction of protocols to suppress spin dephasing  and in the development of schemes for imprinting electron spin states on nuclear memory. 

 We thank Leo DiCarlo, Alex Johnson, and Edward Laird  for technical
 contributions. We thank Bill Coish, Frank Koppens, Daniel Loss, and Amir Yacoby for useful discussion. This work was supported by ARO/IARPA, DARPA, NSF-NIRT (EIA-0210736), the Harvard Center for Nanoscale Systems, and a Pappalardo fellowship (JMT).  Research at UCSB supported in part by QuEST, an NSF Center.


\begin{thebibliography}{45}
\bibitem[{foo({\natexlab{a}})}]{foot_funnel}
\bibinfo{journal}{Although the true functional form relating the position of
  the resonance to $\Bo$ is a power-law, a linear dependence fits the
  data well over the range of polarization accessed in this experiment.}
  {\natexlab{a}}
\bibitem[{foo({\natexlab{e}})}]{foot_recal}
\bibinfo{journal}{A re-calibration is required to account for a slight shift in
  the position of the resonance during pausing in (1,1).}  {\natexlab{e}}
\end{thebibliography}
\end{document}